\documentclass[aps,prx,amsfonts,floatfix,superscriptaddress,longbibliography,twocolumn,footinbib]{revtex4-2}
\usepackage[utf8]{inputenc}
\usepackage{amsmath}    
\usepackage{amssymb}
\usepackage{mathbbol}
\usepackage{graphicx}   
\usepackage{verbatim}   
\usepackage{subfigure}  
\usepackage{hyperref} 
\usepackage{scalerel}
\usepackage{cancel}
\usepackage{soul}
\usepackage[normalem]{ulem}
\usepackage[dvipsnames]{xcolor}
\usepackage{siunitx}
\usepackage{physics}
\usepackage{enumerate} 
\usepackage{empheq}

\begin{document}

\title{Proposal for many-body quantum chaos detection with single-site measurements}

\author{Isa\'ias Vallejo-Fabila}
\affiliation{Department of Physics, University of Connecticut, Storrs, Connecticut 06269, USA}

\author{Adway Kumar Das}
\affiliation{Department of Physical Sciences, Indian Institute of Science Education and Research Kolkata, Mohanpur 741246, India}

\author{Sayan Choudhury}
\affiliation{Harish-Chandra Research Institute, a CI of Homi Bhabha National Institute, Chhatnag Road, Jhunsi, Allahabad 211019, India}

\author{Lea F. Santos}
\affiliation{Department of Physics, University of Connecticut, Storrs, Connecticut 06269, USA}

\begin{abstract}
We demonstrate that the long-time dynamics of an observable associated with a single lattice site is sufficient to determine whether a many-body quantum system exhibits level statistics characteristic of random matrix theory, a widely used diagnostic of quantum chaos. In particular, we focus on the partial survival probability and spin autocorrelation function at a single site, both evolved under a disordered spin-1/2 chain, which is a setup realizable in current experimental platforms. Given the precision and timescales currently achievable, our results indicate that the detection of many-body quantum chaos is feasible, but constrained to small system sizes.
\end{abstract}

\maketitle

In isolated many-body quantum systems, many-body quantum chaos -- characterized by highly delocalized (ergodic) eigenstates and strongly correlated energy levels consistent with random matrix theory -- provides the mechanism that makes thermalization~\cite{Srednicki1994,ZelevinskyRep1996,Borgonovi2016,Alessio2016,Deutsch2018} possible without any coupling to an external bath. Beyond thermalization, many-body quantum chaos is crucial for understanding the scrambling of quantum information, where initially localized information rapidly spreads throughout the Hilbert space~\cite{Chavez2019,Pilatowsky2020,Xu2020,Xu2024,Swingle2016,Hosur2016,Bentsen2019,Vikram2024}. It has also played a key role in the study of black holes~\cite{Sekino2008,Cotler2017,Magan2018}, leading to the derivation of a universal bound on chaos~\cite{Maldacena2016JHEP,Murthy2019}, and it has strong connections with quantum computing~\cite{Flambaum2000A,Georgeot2000,Georgeot2000b,Berman2001} and entanglement generation~\cite{SantosEscobar2004,Karthik2007,Brown2008,Bertini2019,DeTomasi2020,Dowling2024}, relationships that have seen increasing interest in recent years~\cite{Cotler2023,Bhore2023,Choi2023,Ippoliti2022,Ippoliti2023,Mark2024}.

Excellent agreement has been observed between the level statistics predicted by random matrix theory and experimental spectral data from complex systems, such as heavy nuclei~\cite{Haq1982} and atoms~\cite{Flambaum1998,Maier2015}. In contrast, direct access to the full spectrum is more limited in quantum computers and platforms based on interacting ions and cold atoms, where the main interest is typically quantum dynamics.

In this work, we show that spectral correlations as in random matrix theory can be identified in many-body quantum systems by analyzing their long-time dynamics. In contrast to Ref.~\cite{Das2025}, where the proposal for the detection of dynamical manifestations of spectral correlations relied on measurements across the entire chain, here we demonstrate that it is sufficient to monitor a single site. 

In chaotic systems, the survival (or return) probability, defined as the fidelity between an initial state and its time-evolved counterpart,   exhibits a characteristic dip below its saturation value, referred to as the correlation hole (or ramp) \cite{Leviandier1986,Alhassid1992,Guhr1990,Torres2017,Torres2017Philo,Das2023a,Das2022a,Das2022,Das2025arxivBRM}.  This feature does not emerge in systems with uncorrelated eigenvalues. Experimentally, however, observing the correlation hole is challenging, because in many-body quantum systems, it occurs at long times that grow exponentially with system size~\cite{Schiulaz2019}, and for values of the survival probability that are very small, requiring precision beyond current experimental capabilities. Moreover, the survival probability is a nonlocal quantity, introducing additional challenges for measurement.

An experimental approach to address the issue of nonlocality was recently introduced in~\cite{Karch2025}, where a partial survival probability was measured by restricting observations of the evolved state to a portion of the system, which the authors termed the ``subsystem Loschmidt echo''. Nevertheless, this method does not overcome the fundamental obstacle of long timescales required for the emergence of the correlation hole, since this timescale is determined by the spectrum, specifically by  the spacings between eigenvalues, which remain unaffected by partial projective measurements.

The timescales and precision currently achievable in experiments can be matched by working with small chains~\cite{Das2025,Dong2025}. In Ref.~\cite{Das2025}, it was shown that the correlation hole in the survival probability can be observed in chaotic many-body quantum systems with as few as six sites. Building on this approach, we maintain a small chain size and further address the challenge of nonlocality by focusing on measurements at a few sites, and even down to a single site.

We investigate a chain of spin-1/2 particles with nearest-neighbor couplings and onsite disorder, a setup accessible to existing experimental platforms~\cite{Wei2018,Peng2023,Hild2014,Signoles2021,Smith2016,Dong2025} and quantum computers~\cite{Smith2019,Zhu2021}. We show that both the partial survival probability~\cite{Karch2025} and the spin autocorrelation function~\cite{Joshi2022Science} exhibit a correlation hole when the system is chaotic. For both observables, consistent with current experimental capabilities, the correlation hole could be detected using chains with six sites and measurements on a single site. This is certainly the case for the spin autocorrelation functions, while for the survival probability, the curves are smoother and the correlation hole is better visible if one measures all six sites or the four central sites, excluding the edges.

{\em Model and initial state.--} We study a one-dimensional isotropic spin-1/2 Heisenberg system with onsite disorder, homogeneous nearest-neighbor couplings, and open boundary conditions. Disordered spin-1/2 models have been experimentally implemented across a variety of platforms, including nuclear magnetic resonance  systems~\cite{Wei2018,Peng2023}, ultracold atoms~\cite{Hild2014}, Rydberg atom arrays~\cite{Signoles2021}, and trapped ions~\cite{Smith2016}. 
The Hamiltonian considered here is given by
\begin{align}
	\label{eq_MBL}
	\hat{H} =  \hat{H}_z &+ \frac{J}{4} \sum_{k = 1}^{L-1} \left(\hat{\sigma}_k^x\hat{\sigma}_{k+1}^x + \hat{\sigma}_k^y\hat{\sigma}_{k+1}^y \right) , \\
     \hat{H}_z &=  \sum_{k=1}^L \frac{h_k }{2}\hat{\sigma}_k^z + \frac{J}{4} \sum_{k=1}^{L-1} \hat{\sigma}_k^z\hat{\sigma}_{k+1}^z  \nonumber ,
\end{align}
where $L$ is the number of sites, $\hat{\sigma}^x_k, \hat{\sigma}^y_k, \hat{\sigma}^z_k$ are the Pauli matrices acting on the $k$th site, and $J=1$ sets the energy scale. The random onsite fields $h_k$ are uniformly sampled from the interval $[-h, h]$. This Hamiltonian conserves the total magnetization along the $z$-direction, ${\cal S}_z = \sum_k \hat{\sigma}_k^z/2$, leading to a block-diagonal structure where the full Hilbert space decomposes into $L+1$  independent sectors. We consider chains with even $L$ and focus on the largest sector with ${\cal S}_z =0$, which has dimension $D=L!/(L/2)!^2$. 

Hamiltonian~(\ref{eq_MBL}) has played a central role in investigations of many-body localization~\cite{SantosEscobar2004,Pal2010,Torres2015,Nandkishore2015,Luitz2017,Suntajs2020,Sierant2025}. When the disorder strength is comparable to the interaction strength ($h\sim J$), the system enters a chaotic regime characterized by level repulsion and spectral correlations as in random matrix theory~\cite{Avishai2002,Santos2009JMP,Das2019,Das2023}. For strong disorder ($h\gg J$), level statistics in finite systems cross over to Poissonian behavior, indicating the onset of localization. Our focus is on the chaotic regime, so we use $h=0.5$ [see figures for level repulsion in the Supplemental Material (SM)].

To probe the system's dynamics, we initialize it in an eigenstate of the unperturbed Hamiltonian $\hat{H}_z$.  We select initial states $|\Psi(0)\rangle$ with energies that lie near the center of the spectrum, $|\langle \Psi(0)|\hat{H}|\Psi(0)\rangle\sim 0$, where the density of states is high and strong eigenstate hybridization facilitates the development of chaotic behavior~\cite{ZelevinskyRep1996,Borgonovi2016}. The following notation is used for the eigenvalues and eigenstates of $\hat{H}$ and $\hat{H}_z$, respectively, $\hat{H}|\alpha\rangle = E_{\alpha} |\alpha\rangle$ and $\hat{H}_z|n\rangle = \epsilon_{n} |n\rangle$.

We perform averages over $10^{4}$ samples, from which $0.1D$ are initial states in the middle of the spectrum and $10^{4}/0.1D$ correspond to disorder realizations. An additional running average is also performed to further smooth the curves. Averages are needed, because the survival probability and the spin autocorrelation function at long times are non-self-averaging~\cite{Prange1997,Schiulaz2020,Torres2020,Vallejo2024}. If averages are not performed, the correlation hole may be hidden by fluctuations. 

{\em Correlation hole and survival probability.--} The presence of spectral correlations in a quantum system can be investigated either by directly analyzing the energy eigenvalues, using quantities such as the level spacing distribution and the level number variance~\cite{MehtaBook,Guhr1998}, or by examining Fourier transforms of the energy spectrum, as in the case of the spectral form factor~\cite{MehtaBook},
\begin{align}
	\label{eq_SFF}
	S_{FF}(t) &= \dfrac{1}{D^2} \left\langle \sum_{\alpha, \beta}e^{-i(E_\alpha - E_\beta)t} \right\rangle,
\end{align}
where $\langle . \rangle$ indicates average over initial states and disorder realizations. In chaotic models, the spectrum contains a disconnected part associated with the density of states of the system and a connected part, whose Fourier transform leads to the two-level form factor~\cite{MehtaBook, Das2025arxivBRM} responsible for the correlation hole. This signature of quantum chaos can also be captured by the partial spectral form factor~\cite{Joshi2022,Yoshimura2025},  which is a generalization of $S_{FF}(t)$ for a subsystem of the total many-body system. Similarly to the   survival probability, the partial $S_{FF}(t)$ encompasses information about the spectral correlations and the eigenstates~\cite{Dong2025,Fritzsch2025}.

To detect the correlation hole dynamically after a quench from the initial Hamiltonian $\hat{H}_z$ to the final Hamiltonian $\hat{H}$, one can resort to the survival probability,
\begin{equation}
	\label{eq_SP_def}
		S_P(t) = \left| \langle \Psi(0)|\Psi(t)\rangle \right|^2  =  \sum_{\alpha,\beta} |C^{n_0}_\alpha|^2 |C^{n_0}_\beta|^2 e^{-i(E_\alpha - E_\beta)t}  ,
\end{equation}
where $C^{n_0}_\alpha = \langle \alpha|\Psi(0)\rangle$ are the components of the initial state in the energy eigenbasis and $|\Psi(0)\rangle = |n_0 \rangle$ represents one of the eigenstates of $\hat{H}_z$. The survival probability is a nonlocal quantity that contains both information about the initial state and the spectrum. The components $C^{n_0}_\alpha$ and the disconnected part of the spectrum control the initial decay of $S_P(t)$ \cite{Khalfin1958,Flambaum2001b,Tavora2017}. The connected part of the spectrum gets manifested at long times, after the dynamics resolves the discreteness of the spectrum.

At long times, the evolution of the survival probability in strongly chaotic systems develops a correlation hole  before saturating at its asymptotic value $\overline{S_P} = \sum_{\alpha} |C^{n_0}_\alpha|^4$. In the model described by Eq.~(\ref{eq_MBL}), the ramp towards saturation begins at a time that scales as $D^{2/3}$ and for any chaotic system, including those modeled by random matrices, saturation of $S_P(t)$ occurs at the Heisenberg time, which scales as $D$ \cite{Schiulaz2019}. Therefore, both timescales grow exponentially with system size, making it experimentally challenging to access the full temporal range of the correlation hole, even for relatively small system sizes. 

A further challenge lies in the small values that the survival probability takes in the correlation hole region. For initial states near the middle of the spectrum, both the asymptotic value $\overline{S_P}$, which quantifies the number of energy eigenstates participating in the dynamics, and the minimum value of the survival probability at the correlation hole, $(S_P)_\text{min}$, approach the predictions for random matrices from Gaussian orthogonal ensembles, namely $\overline{S_P^{\text{GOE}}} \approx 3/D$ and $(S^{\text{GOE}}_P)_\text{min} \approx 2/D$, respectively~\cite{Alhassid1992,Schiulaz2019}. These values are small even for chains of just 10 sites. Consequently, the combination of long-time evolution, low values of $S_P(t)$, and the need to measure a nonlocal observable makes experimental detection of the correlation hole in $S_P(t)$ particularly demanding. We discuss next how these challenges can be overcome.

{\em Partial survival probability.--} The partial survival probability is the survival probability of a subsystem of size $L_s\leq L$ \cite{Karch2025}. For example, consider the initial state $ |\Psi(0) \rangle = \ket{ \uparrow \downarrow \uparrow \downarrow } $ evolving under the Hamiltonian in Eq.~(\ref{eq_MBL}) with ${\cal S}_z =0$ and $L=4$. The partial survival probability, $S_P^{(L_s,L)}(t)$, for two sites ($L_s=2$) at the center of the chain is given by
\begin{align} 
   S_P^{(2,4)}(t) &= \left| \langle \uparrow \boxed{\downarrow \uparrow} \downarrow|e^{-i H t} |\uparrow \boxed{\downarrow \uparrow} \downarrow \rangle \right|^2 \nonumber \\
   &+ \left| \langle \downarrow \boxed{\downarrow \uparrow} \uparrow|e^{-i H t} |\uparrow \boxed{\downarrow \uparrow} \downarrow \rangle \right|^2 .
\end{align}
The partial survival probability contains the survival probability (first term in the equation above) and additional contributions  from multiple configurations of the discarded subsystem, which rise the saturation value $\overline{S_P^{(L_s,L)}}$ above $\overline{S_P}$.

In Fig.~\ref{fig:SP}, we show the partial survival probability for a chain with $L=6$ sites [Figs.~\ref{fig:SP}(a)-(c)] and a chain with $L=12$ sites [Figs.~\ref{fig:SP}(d)-(f)]. The focus is on long times. The horizontal dashed line indicates the saturation value of $S_P^{(L_s,L)}(t)$, and the ramp below this line going up to $\overline{S_P^{(L_s,L)}}$ characterizes the correlation hole. 

\begin{figure}[h]
\centering
\includegraphics[scale=0.335]{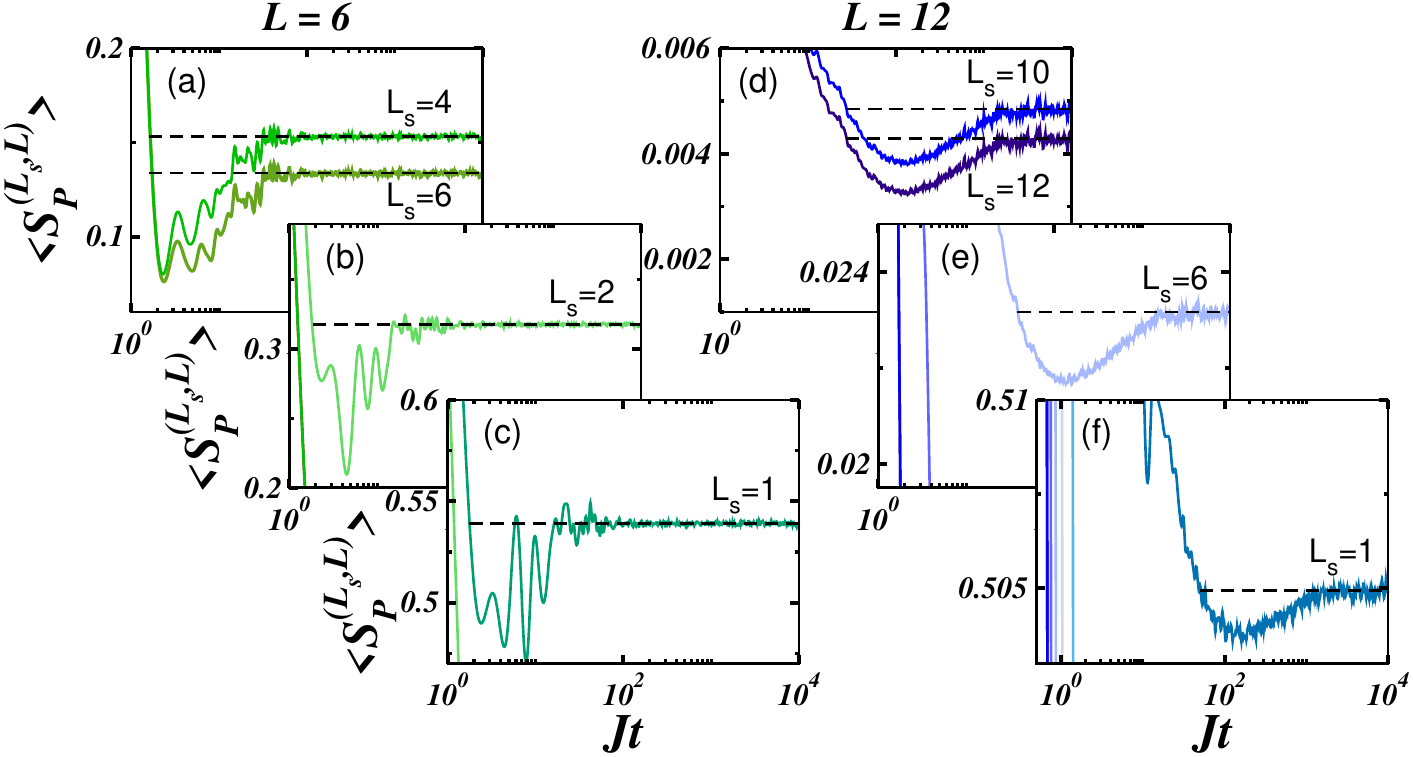}
\caption{Partial survival probability for a chain with (a)-(c) $L=6$ and (d)-(f) $L=12$. The horizontal dashed line indicates the saturation value of $S_P^{(L_s,L)}(t)$. The measurement is performed over a subsystem of size $L_s$, as indicated in the panels. 
}
\label{fig:SP}
\end{figure}

The timescale for the emergence of the correlation hole for $L=12$ [Figs.~\ref{fig:SP}(d)-(f)] is too long and exceeds what is currently accessible in experiments. Furthermore, as evident in Figs.~\ref{fig:SP}(d)-(f), this timescale does not change as $L_s$ decreases. The spectrum characterizing the system, and thus the timescale for the emergence of dynamical manifestations of spectral correlations, is the same, independently of the number of sites in which the measurement is performed. What changes as $L_s$ decreases is the minimum point of $S_P^{(L_s,L)}(t)$ and its saturation value, both of which increase [compare the scales of the $y$-axis from Fig.~\ref{fig:SP}(d) to Fig.~\ref{fig:SP}(f) and see additional figures in the SM]. 

The results for the chain with $L=6$ in Figs.~\ref{fig:SP}(a)-(c) are experimentally  promising, since the beginning of the ramp happens for $t<5J^{-1}$ and equilibration takes place for $t\sim10^2 J^{-1}$. Measuring six sites is viable~\cite{Karch2025}, but discarding the sites at the border of the chain and considering only the four central sites is even better in the sense of achieving a larger relative depth [compare the minima and saturation values of the two curves in  Fig.~\ref{fig:SP}(a)]. The correlation hole is also present when the measurement is restricted to the two central sites [Fig.~\ref{fig:SP}(b)] or even a single site [Fig.~\ref{fig:SP}(c)], although the curves exhibit larger fluctuations.

{\em Spin autocorrelation function.--} The spin autocorrelation function at site $k$ is defined as
\begin{align}
	\label{eq_Imb_def}
	A^z_k(t) &= \bra{\Psi(0)} \hat{\sigma}_k^z e^{i \hat{H} t} \hat{\sigma}_k^ze^{-i \hat{H} t} \ket{\Psi(0)},
\end{align}
and its average over $L_s$ sites is given by
\begin{align}
	\label{eq_Imb_ave}
	C^z_{(L_s,L)}(t) &= \frac{1}{L_s}\sum_{k=k_1}^{k_1+L_s-1}A^z_k(t),
\end{align}
where $k_1$ is the first site of the subsystem considered. 
Contrary to the survival probability, this is a local quantity in space, although also nonlocal in time. It has been the subject of several theoretical studies~\cite{Torres2018,Schiulaz2019,Kliczkowski2024} and has been measured experimentally~\cite{Joshi2022Science}. This quantity is related to the imbalance measured in~\cite{Schreiber2015,Bordia2016}.

\begin{figure}[h]
\centering
\includegraphics[scale=0.335]{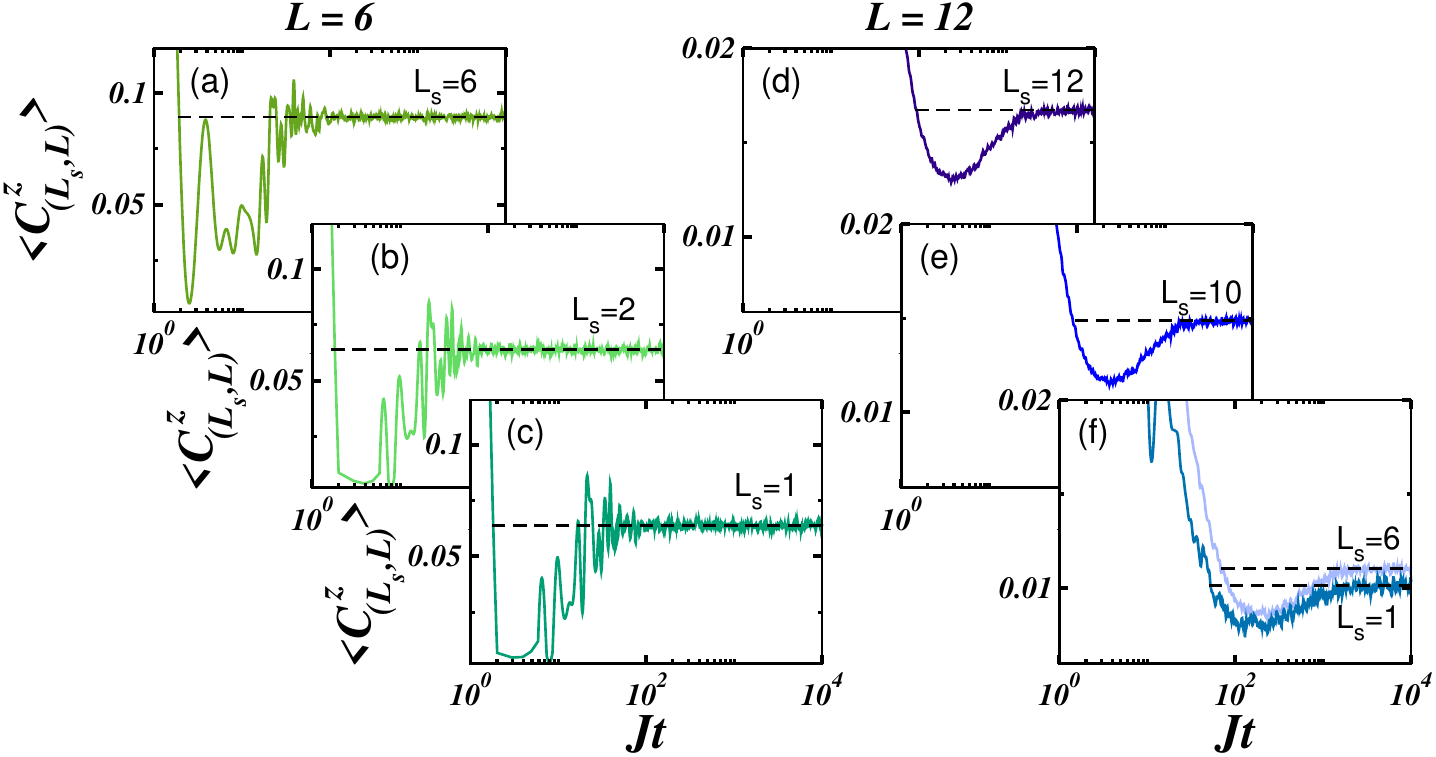} 
\caption{Spin autocorrelation function for a chain with (a)-(c) $L=6$ and (d)-(f) $L=12$. The horizontal dashed line indicates the saturation value of $C^z_{(L_s,L)}(t)$.  The measurement is performed and averaged over $L_s$, as indicated in the panels.
}
\label{fig:Imb}
\end{figure}

In Fig.~\ref{fig:Imb}, we present the spin autocorrelation function for a chain with $L=6$ sites [Figs.~\ref{fig:Imb}(a)-(c)] and a chain with $L=12$ sites [Figs.~\ref{fig:Imb}(d)-(f)], with horizontal dashed lines indicating the saturation values. Since the spin autocorrelation function can take negative values, its fluctuations can be larger than those of the survival probability. For $L=6$, the minimum values reached by $C^z_{(L_s,L)}(t)$ can be smaller than those of $S_P^{(L_s,L)}(t)$, although this trend reverses for $L \geq 8$.

Consistent with the timescale analysis in Fig.~\ref{fig:SP}, the results in  Figs.~\ref{fig:Imb}(a)-(c) indicate that the correlation hole in the spin autocorrelation function can be detected in a chain with only six sites. Crucially, this observable offers a key advantage over the survival probability: the number of sites $L_s$ involved in the measurement of $C^z_{(L_s,L)}(t)$ does not significantly affect the results. Thus, the correlation hole in $C^z_{(L_s,L)}(t)$ can be revealed with measurements on a single site as effectively as with many sites. 

{\em Comparison between quantities.--} A better idea of the differences and similarities between the spin autocorrelation function and survival probability can be understood by writing $A_k^z(t)$ in terms of $S_P(t)$ as 
\begin{align}
	\label{eq_Imb_SP}
  \langle A^z_k(t)  \rangle &= \langle S_P(t)\rangle + \langle P_k^{n_0'}(t)\rangle + \langle P_k^n(t)\rangle , \\
 \langle P_k^{n_0'}(t)\rangle &= \left\langle \chi_k^{n_0} \chi_k^{n_0'} \sum_{\alpha, \beta} C_{\alpha} ^{n_0} C_{\beta} ^{n_0} C_{\alpha} ^{n_0'} C_{\beta} ^{n_0'} e^{-i(E_{\alpha} - E_{\beta})t} \right\rangle \nonumber \\
  \langle P_k^n(t)\rangle&=\left\langle\sum_{n \neq n_0,n_0' } \!\!\!\!\chi_k^{n_0} \chi_k^{n} \sum_{\alpha, \beta} C_{\alpha} ^{n_0} C_{\beta} ^{n_0} C_{\alpha} ^{n} C_{\beta} ^{n} e^{-i(E_{\alpha} - E_{\beta})t} \right\rangle \nonumber, 
\end{align}
where $\chi_k^{n} = \langle n|\sigma_k^z|n\rangle$ and $|n_0'  \rangle =\sigma_1^x \sigma_2^x \ldots \sigma_L^x |n_0\rangle$ corresponds to the state obtained by rotating the spins of the initial state $|n_0  \rangle$ by an angle $\pi$ around the $x$-direction. The term $\langle P_k^{n_0'}(t)\rangle$ involving the state $|n_0'  \rangle$ is negative at long times. Furthermore, there is an interesting interplay between the survival probability and $\langle P_k^n(t)\rangle$ in Eq.~(\ref{eq_Imb_SP}), as discussed next. 

\begin{figure}[h]
\centering
\vskip 0.2 cm
\includegraphics[scale=0.45]{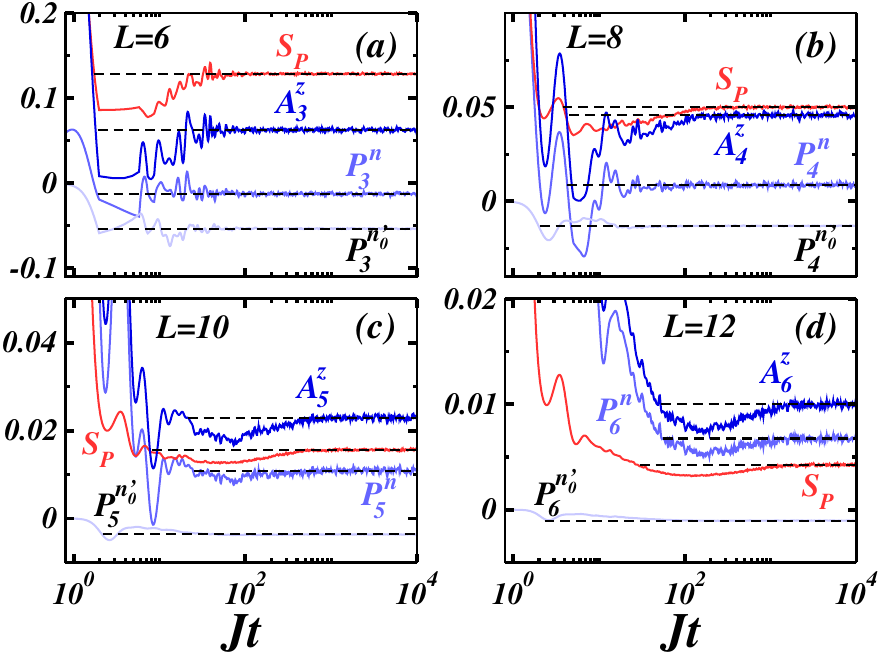} 
\caption{Comparison between the evolution of the spin autocorrelation function on the central site, $\langle A^z_{L/2}(t) \rangle$, with the evolution of the survival probability $\langle S_P(t) \rangle$ and the two other terms in Eq.~(\ref{eq_Imb_SP}): $\langle P_{L/2}^{n_0'}(t)\rangle$ and $\langle P_{L/2}^n(t)\rangle $. 
} 
\label{fig03}
\end{figure}

For long times and small chains, $L=6$ [Fig.~\ref{fig03}(a)] and $L=8$ [Fig.~\ref{fig03}(b)], the hierarchy $|\langle P_{n}(t)\rangle| < |\langle P_k^{n_0'}(t)\rangle| < \langle S_P(t)\rangle$ holds and the survival probability is the main contributor to the spin autocorrelation function~\footnote{For these two system sizes, we excluded from the curves few time points at which $\langle A^z_{k}(t)\rangle$ is negative}. In this case,   $\langle A^z_{k}(t)\rangle$ is slightly smaller than $\langle S_P(t)\rangle$.  As the system size increases, the contributions  from $\langle P_{n}(t)\rangle$ are enhanced and those from the survival probability diminish. For $L=10$ [Fig.~\ref{fig03}(c)], $|\langle P_{n}(t)\rangle|$ surpasses $|\langle P_k^{n_0'}(t)\rangle|$, and $\langle A^z_{k}(t)\rangle$ becomes slightly larger than $\langle S_P(t)\rangle$. For $L\geq12$ [Fig.~\ref{fig03}(d)], the long-time contributions from $\langle P_{n}(t)\rangle$ to $\langle A^z_k(t)\rangle$ exceed those from $\langle S_P(t)\rangle$. 

This change with system size of the dominant term in $\langle A^z_k(t)\rangle$ also impacts the visibility of the correlation hole. While the survival probability retains a nearly constant relative depth as $L$ increases, approaching the random matrix prediction $[\overline{S_P^{\text{GOE}}} - (S_P^{\text{GOE}})_\text{min})]/\overline{S_P^{\text{GOE}}} = 1/3$, the relative depth of the correlation hole in the spin autocorrelation function progressively decreases with $L$ and eventually vanishes in the thermodynamic limit~\cite{Lezama2021}.

As a final remark, we note that the fluctuations observed in the correlation hole for chains with six sites ($D = 20$) can be reduced by  adding random couplings. 
However, even in the extreme case of full random matrices, the early part of the ramp for $D = 20$ still exhibits noticeable fluctuations. This happens because at the timescale of the ramp and for such small matrices, there is  overlap between the effects of the connected and disconnected parts of the spectrum (see figures in the SM). In contrast, for large matrices, the dynamical behavior associated with the shape of the energy distribution of the initial state, such as power-law decays of the survival probability~\cite{Schiulaz2019,Tavora2017}, emerge at intermediate times, while the ramp, caused by the correlated eigenvalues, appears at long times~\cite{Schiulaz2019}. In small matrices, this separation is less distinct, especially at the point where the ramp begins. 

{\em Conclusions.--} This work demonstrates that the long-time dynamics of the partial survival probability and spin autocorrelation function are effective probes of many-body quantum chaos, as both exhibit the correlation hole (ramp). Remarkably, this signature of spectral correlations is detectable with single-site measurements in small disordered spin chains accessible to current experiments. Despite the inherent challenges posed by long timescales and low signal amplitudes, our results show that the experimental detection of dynamical fingerprints of many-body quantum chaos are within reach of existing technology.

{\em Acknowledgments.--}
 The authors thank Anandamohan Ghosh for useful discussions and acknowledge support from the National Science
Foundation Center for Quantum Dynamics on Modular Quantum Devices (CQD-MQD) under Award Number
2124511. A.~K.~D.~is supported by an INSPIRE fellowship, DST, India. S.~C. thanks DST, India for support through SERB project SRG/2023/002730.


\bibliography{biblio2025,partial_SP_Adway}


\newpage


\title{Supplemental material for EPAPS\\
Entropy of isolated quantum systems after a quench}

\author{Lea F. Santos}
\affiliation{Department of Physics, Yeshiva University, New York, NY 10016, USA}
\author{Anatoli Polkovnikov}
\affiliation{Department of Physics, Boston University, Boston, MA 02215, USA}
\author{Marcos Rigol}
\affiliation{Department of Physics, Georgetown University, Washington, DC 20057, USA}
\affiliation{Kavli Institute for Theoretical Physics, University of California, Santa Barbara,
California 93106, USA}

\maketitle


\onecolumngrid

\vspace*{0.4cm}

\begin{center}

{\large \bf Supplemental Material: 
\\ Proposal for many-body quantum chaos detection with single-site measurements}\\

\vspace{0.6cm}

 Isa\'ias Vallejo-Fabila$^1$, Adway Kumar Das$^2$, Sayan Choudhury$^3$, and Lea F. Santos$^1$,\\

$^1${\it Department of Physics, University of Connecticut, Storrs, Connecticut 06269, USA}

$^2${\it Department of Physical Sciences, Indian Institute of Science Education and Research Kolkata, Mohanpur 741246, India}

$^3${\it Harish-Chandra Research Institute, a CI of Homi Bhabha National Institute, Chhatnag Road, Jhunsi, Allahabad 211019, India}

\end{center}

\vspace{0.6cm}

\twocolumngrid

This supplemental material provides additional figures supporting the discussions in the main text. Section~\ref{Sec:DOS} presents the density of states and level statistics for the spin-1/2 model defined in Eq.~(1) of the main text. In Sec.~\ref{Sec:Relative}, we show how the relative depth for the onset of the correlation hole  scales with the number of measured sites, $L_s$.  Section~\ref{Sec:Random} examines how introducing random couplings helps to smooth the curves of the partial survival probability and spin autocorrelation function for a chain with only six sites.

\section{Density of states and level statistics}
\label{Sec:DOS}

\begin{figure}[ht]
\centering
\vskip 0.2 cm
\includegraphics[scale=0.32]{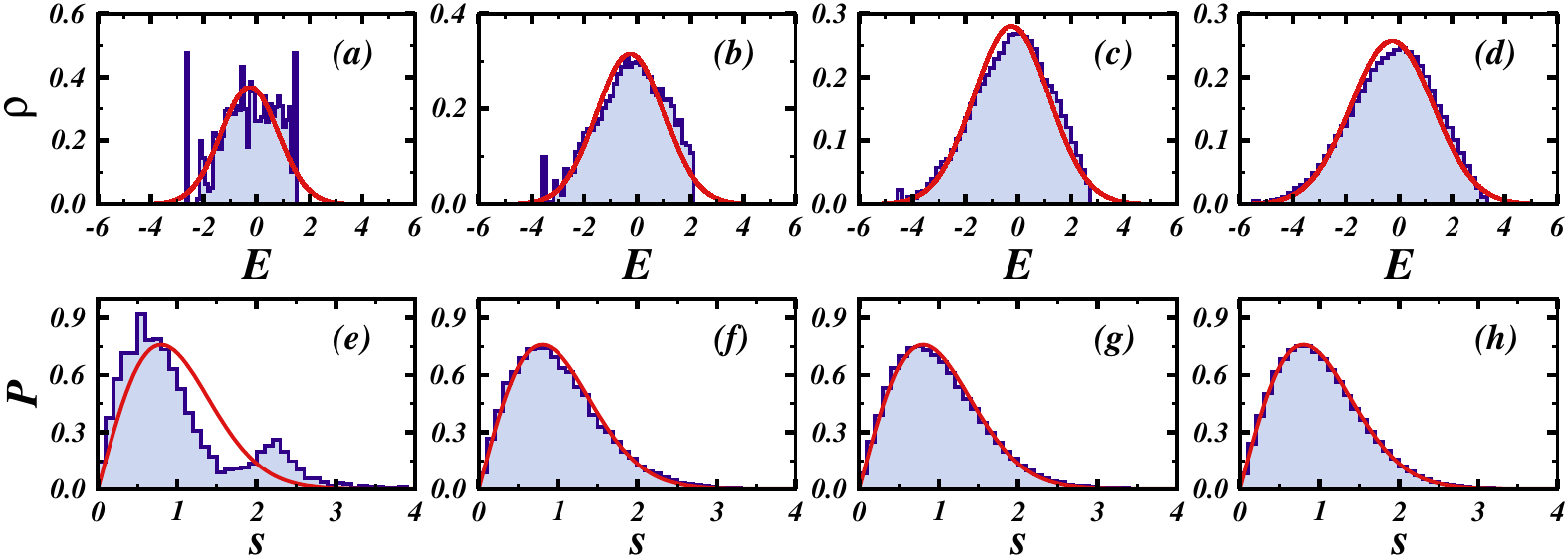} 
\caption{(a)-(d) Density of states and (e)-(h) level spacing distribution for (a),(e) $L=6$; (b),(f) $L=8$; (c),(g) $L=10$; and (d),(h) $L=12$. Averages are performed over disorder realizations. Red solid line indicates (a)-(d) Gaussian distribution and (e)-(h) Wigner-Dyson distribution. The model is the same as in Eq.~(1) of the main text with $J=1$ and $h=0.5$.
} 
\label{fig:SM01}
\end{figure}

Figure~\ref{fig:SM01} presents the density of states, $\rho(E) = \sum_{\alpha} \delta(E-E_{\alpha})$, 
[Fig.~\ref{fig:SM01}(a)-(d)] and the  distribution $P$ of spacings $s$ between unfolded energy levels [Fig.~\ref{fig:SM01}(e)-(h)] for the chaotic spin-1/2 system in the Eq.~(1) of the main text. The system size grows from left to right, ranging from $L=6$ [Fig.~\ref{fig:SM01}(a),(e)] to $L=12$ [Fig.~\ref{fig:SM01}(d),(h)]. 

The density of states follows a Gaussian shape [red solid line in Fig.~\ref{fig:SM01}(a)-(d)] and the level spacing distribution exhibits the Wigner-Dyson distribution, $P(s) = (\pi s/2) \exp(-\pi s^2/4)$ [red solid line in Fig.~\ref{fig:SM01}(e)-(h)] characteristic of quantum chaotic systems. For $L=6$, where the system has only three spins pointing up in the $z$-direction, deviations from typical many-body behavior emerge~\cite{Schiulaz2018,Zisling2021}, which contribute to the fluctuations observed in the ramp of the correlation hole in Figs.~1-2 of the main text. Nonetheless, level repulsion is already evident at this small system size, which is also reflected in the dynamics, as seen in those same Figs.~1-2 of the main text.

\section{Relative depth and timescale for the ramp}
\label{Sec:Relative}

\begin{figure}[h]
\centering
\vskip 0.2 cm
\includegraphics[scale=0.55]{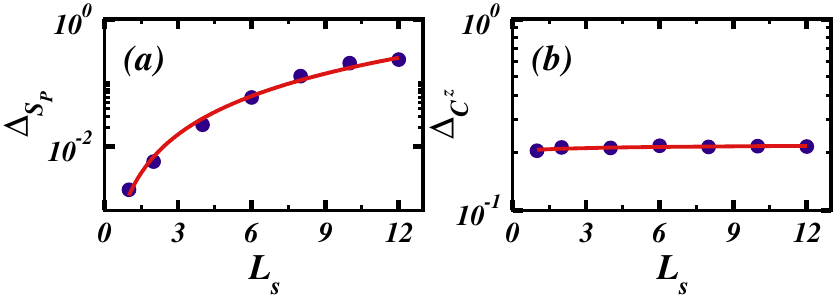} 
\caption{Relative depth for the (a) partial survival probability and (b) spin autocorrelation function vs the number of measured sites $L_s$. The chain size is $L=12$ and $\Delta_{S_P} \propto L_s^2$. The model is the same as in Eq.~(1) of the main text with $J=1$ and $h=0.5$.
} 
\label{fig:SM02}
\end{figure}

Figure~\ref{fig:SM02} show the relative depth,
\begin{equation}
    \Delta_O = \frac{\overline{O} - O_{\text{min}}}{\overline{O}}
\end{equation}
for the partial survival probability, $O=S_P^{(L_s,L)}$, [Fig.~\ref{fig:SM02}(a)] and the spin autocorrelation function, $O=C^z_{(L_s,L)}$, [Fig.~\ref{fig:SM02}(b)] as a function of the number $L_s$ of measured sites. A chain with a total of $L=12$ sites is considered. 

As $L_s$ decreases, both the minimum value of $S_P^{(L_s,L)}$ and its saturation value increase [see Fig.~1(d)-(f) in the main text], which could be experimentally advantageous. However, as seen in Fig.~\ref{fig:SM02}(a), the relative depth decreases as $L_s$ decreases following the fitting $\Delta_{S_P} \propto L_s^2$, thus requiring good experimental precision to distinguish $(S_P^{(L_s,L)})_{\text{min}}$ from $\overline{S_P^{(L_s,L)}}$.

In contrast to the partial survival probability, the relative depth for the spin autocorrelation function in Fig.~\ref{fig:SM02}(b) is independent of $L_s$. This emphasizes what we wrote in the main text, that the correlation hole can be revealed with measurements
on a single site as effectively as with measurements on many sites.

\section{Random couplings}
\label{Sec:Random}

\begin{figure}[h]
\centering
\vskip 0.2 cm
\includegraphics[scale=0.45]{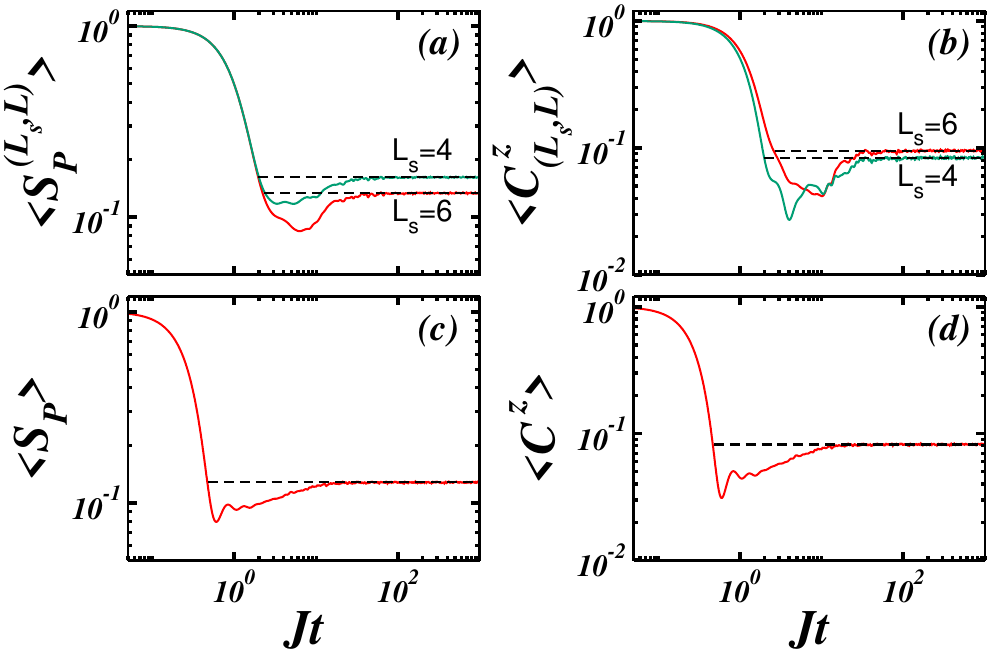} 
\caption{(a), (c) Survival probability $ \langle S_P^{(L_s,L)}(t) \rangle$ and (b), (d) site-averaged spin autocorrelation function $ \langle C^z_{(L_s,L)}(t) \rangle$ for the (a)-(b) model with random couplings in Eq.~(\ref{Eq:HamSM}) with six sites and $L_s=4, 6$ and for (c)-(d) GOE random matrices of dimension $D=20$ and $L_s=L$. 
} 
\label{fig:SM03}
\end{figure}

By adding random couplings to the spin-1/2 system with onsite disorder, it is possible to further smooth the curves for the survival probability and spin autocorrelation function for chains with six sites. In Fig.~\ref{fig:SM03}, we show results for the (partial) survival probability [Fig.~\ref{fig:SM03}(a)] and site-averaged spin autocorrelation function [Fig.~\ref{fig:SM03}(b)] for the following Hamiltonian, 
\begin{align}
	\label{Eq:HamSM}
	\hat{H} =  \sum_{k=1}^L \frac{h_k }{2}\hat{\sigma}_k^z  &+  \sum_{k=1}^{L-1} \frac{J'}{4}\left(\hat{\sigma}_k^x\hat{\sigma}_{k+1}^x + \hat{\sigma}_k^y\hat{\sigma}_{k+1}^y + \hat{\sigma}_k^z\hat{\sigma}_{k+1}^z \right)  ,
\end{align}
where $J' = 1 + \delta J$, with $\delta J \in [-0.5,0.5]$ being uniform random numbers chosen for each random realization and, as before, $h=0.5$. The curves are significantly smoother than those in the main text for both $L_s=6$ and $L_s=4$.

In Fig.~\ref{fig:SM03}, we also show results for the survival probability, $\langle S_P(t) \rangle = \langle S_P^{(L,L)}(t) \rangle$ [Fig.~\ref{fig:SM03}(c)] and site-averaged spin autocorrelation function $\langle C^z(t) \rangle = \langle C^z_{(L,L)}(t) \rangle$ [Fig.~\ref{fig:SM03}(d)] evolved under full random matrices from a Gaussian orthogonal ensemble (GOE). The matrices have dimension $D=20$ as the dimension considered for the spin-1/2 model with $L=6$. This is an unphysical exercise, since there are no basis associated with random matrices. The purpose is simply to show that even in this case, if the dimension is small, there are oscillations at the beginning of the ramp caused by overlaps between the effects of one-level (density of states) and two-level correlation functions.


\end{document}